# X-ray absorption spectroscopy measurement on the LaO$_{1-x}$F$_x$FeAs system

A. Ignatov[1], C. L. Zhang[1], M. Vannucci[1], M. Croft[1,2] T. A. Tyson[1,3], D. Kwok[1], Z. Qin[3], S.-W. Cheong[1]

[1]Department of Physics and Astronomy & Rutgers Center for Emergent Materials, Rutgers University, Piscataway, NJ 08854 and
[2]Brookhaven National Laboratory, Upton, NY 11973
[3] Department of Physics, New Jersey Institute of Technology, 161 Warren Street, University Heights, Newark, New Jersey 07102-1982

**ABSTRACT**

Results of Fe *K*-, As *K*-, and La *L*$_3$-edge x-ray absorption near edge structure (XANES) measurements on LaO$_{1-x}$F$_x$FeAs compounds are presented. The Fe *K*- edge exhibits a chemical shift to lower energy, near edge feature modifications, and pre-edge feature suppression as a result of F substitution for O. The former two changes provide evidence of electron charge transfer to the Fe sites and the latter directly supports the delivery of this charge into the Fe-3*d* orbitals. The As *K*- edge measurements show spectral structures typical of compounds with planes of transition-metal tetrahedrally coordinated to *p*-block elements as is illustrated by comparison to other such materials. The insensitivity of the As-*K* edge to doping, along with the strong Fe-K doping response, is consistent with band structure calculations showing essentially pure Fe-d character near the Fermi energy in these materials. The energy of the continuum resonance feature above the La-*L*$_3$ edge is shown to be quantitatively consistent with the reported La-O inter-atomic separation and with other oxide compounds containing rare earth elements.

## I. Introduction

The discovery of Fe-based pnictide, and now chalcogenide, compounds with high superconducting transition temperatures has captured the attention of the solid-state research community by breaking the high-*T*c cuprate-oxide mold existent for nearly two decades [1-11]. Previously such high temperature superconductors had been limited to perovskite based cuprate compounds with square CuO$_2$ planes in which Cu-O-Cu coordination geometry existed [12-14]. Since the early stages of the high-*T*c research the search for non-cuprate superconductors has been actively but heretofore unsuccessfully pursued. Both early and recent studies of compounds containing similar NiO$_2$ planes are good examples the search for non-cuprate superconductors [14]. A leitmotif of previous high Tc materials was the doping of electrons/holes to the d-orbitals of the CuO$_2$ layers. However, the Cu-O-Cu coordination geometry of the CuO$_2$ layers and the covalent character of Cu-O bond mandated a strong O-character in the doped electrons/holes which supported the superconductivity [13].

Besides transcending the Cu-monopoly, these new Fe-superconductors are not only not oxides but the highest *T*c materials have jumped down two rows in the p-block of the periodic table. Although the Fe atoms do lie in a square planar array their coordination is tetrahedral to As atoms lying above and below the Fe-plane. This opens the possibility of direct Fe interactions without an intervening ligand [1-8]. Therefore, Fe-based high *T*c compounds constitute a truly new class of high *T*c materials to which Cu-O based insights/assumptions can not be applied directly and consideration of alternative mechanisms must be addressed.

"Controlled" electron/hole doping to the *d*-orbitals of transition metals has been the fundamental material tool for varying superconducting, structural, and magnetic properties in perovskite based cuprate [13-14], nickelate [15-16] and manginate [18,19] research. For all above materials, x-ray absorption spectroscopy (XAS) has proved useful in characterizing the "controlled" character of the doping [13-1]. Specifically XAS has been used to track (often decisively) the atomic site and orbital character of the doping induced charge transfer [13, 14, 19]. This work is undertaken to experimentally determine the ground state configuration(s) and the character of doped states in the $LaO_{1-x}F_xFeAs$ superconducting system. Combined analysis of Fe *K*-, As *K*-, and La $L_3$ edges provide information on the unoccupied density of states which are found in good agreement with results of recent band structure calculations [20-22].

## II. Experimental

The $LaO_{1-x}F_xFeAs$, samples were prepared by solid state reaction after the method used by Chen et. al. [3] Specifically mixtures of LaAs, $Fe_2O_3$, Fe (and $LaF_3$ in the doped material) powders were peletized, reacted in Ta crucible, sealed in a quartz ampule under argon atmosphere, and annealed for 50 h at a temperature of 1220 $^o$C. All of the Bragg lines in powder X-ray diffraction measurements could be indexed in the tetragonal ZrCuSiAs-type structure with the space group *P4/nmm* [19]. The x values for the F content cited in the text are those in initial power mixtures. F loss, however, is known to occur in the high temperature annealing process. Kamihara et. al. [1] have studied the superconducting *T*c versus F substitution based upon the diffraction determined F-induced volume contraction and a Vegards Law assumption. Magnetization measurements on our x=0.11 sample showed a superconducting transition at *T*c= 16 K. Using Kamihara et. al.'s [1] results this would correspond to a retained F fraction of $x_K$=0.04 (where the subscript K stands for Kamihara). The x=0.33 specimen had an optimal transition temperature *T*c of 26 K and hence we believe a retained fraction of $x_K$~0.11. As will be seen below, the XANES results for this specimen appeared to continue the F-doping trend established between the x=0 and x=0.11 specimens. It is worth noting that a second x=0 specimen was prepared by a NaCl/KCl flux method described by Nientiedt and Jeitschko [23] and that the diffraction and XAS results for this specimen were essentially the same as for the specimen prepared by solid state synthesis.

The Fe-, As- and Ge-*K*, along with La $L_3$- edge x-ray absorption near edge structure (XANES) measurements were performed in both fluorescence and transmission modes on beamline X-19A at the National Synchrotron Light Source and Brookhaven National Laboratory. Experimental details were discussed elsewhere [14-18]. A focused beam along with a double-crystal Si(111) monochromator was used. All data reported in this paper were collected at room temperature. The Fe *K*- edge energies were calibrated by running a simultaneous standard and have a relative accuracy of better than 0.05 eV. Indeed careful comparison of a large region of the standard spectra were undertaken to optimize the relative energy calibration. The higher-energy harmonics were suppressed by detuning the second crystal of the double crystal monochromator on its rocking curve to 1/2 of maximum intensity at ~200eV above the edge. From two to five scans were collected to ensure reproducibility of the data

To facilitate discussion of the Fe-As hybridization effects Si *K*- edge measurements are reported on similar layered, transition metal silicides with tetrahedral coordination. The Si *K*- edge measurements were performed using InSb (111) crystals in an unfocused beam geometry. The total electron yield mode was uses for the Si *K*- edge measurements to prevent self absorption effects at the low energies involved. Strong absorption prevented use of the simultaneous standard technique so elemental Si samples were run periodically to calibrate energy variations during the Si *K*- edge measurements.

### III. Results and Discussion

*3.1 Fe-K edge results*

Figure 1 shows the near edge Fe-K XAS spectra for LaOFeAs along with those for the standard compounds $Fe^{2+}$ (FeO), $Fe^{3+}$ (LaSrFeO$_4$), and ~$Fe^{4+}$ (SrFeO$_{3-\delta}$) [24]. The ~20 eV region above the steeply rising main edge is dominated by dipole transitions from the core-1*s* states to *p*-states above the Fermi energy. Fe-4*p* final state features in this main edge region ride on top of the step-onset feature associated with transitions to continuum states [14-19, 24,25]. The Fe-4p features splitting can vary strongly due to local bonding/symmetry effects and differing Fe-3*d* state configuration contributions [14-18]. Despite such complicating multiple features, tracking the chemical shift of the main edge to lower energy with decreasing valence is a standard tool for probing variations of formal valence of transition metals [25] (see references 15-19, 24 for Mn, Fe, Co, and Ni examples).

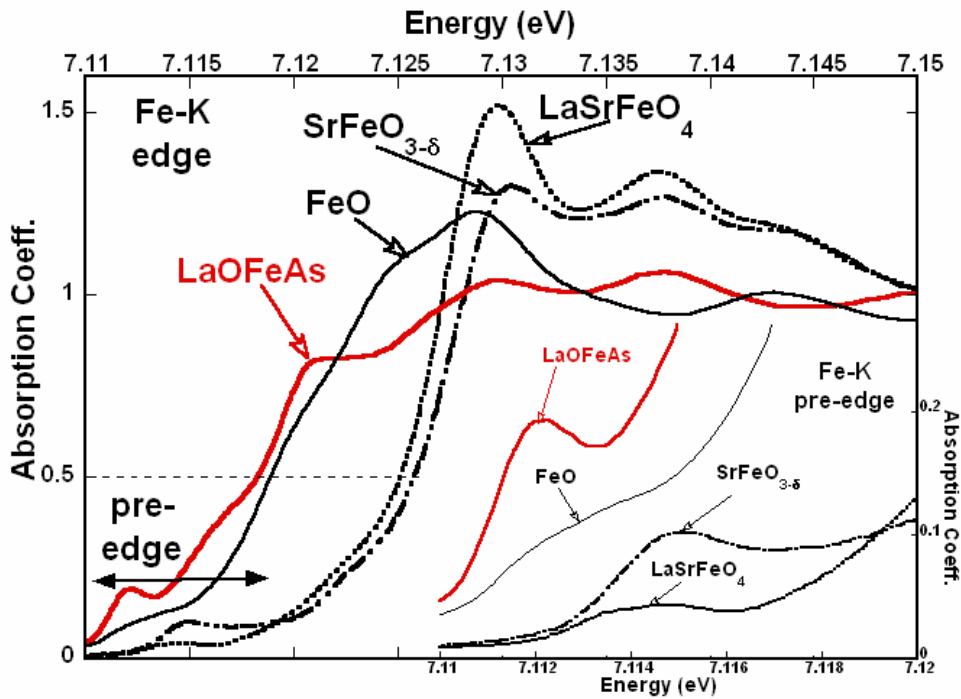

Figure 1: (color on line). Fe-K edge spectra for the standard compounds $Fe^{2+}$ (FeO), $Fe^{3+}$ (LaSrFeO$_4$), and ~$Fe^{4+}$ (SrFeO$_{3-\delta}$) compared to that of LaOFeAs. Inset: An expanded view of the pre-edge region of these spectra.

In Figure 1 a dashed line at the absorption coefficient value of $\mu \sim 0.5$ is included for the purpose of the chemical shift discussion. The chemical shift between the $\sim Fe^{4+}$ ($SrFeO_{3-\delta}$) and $Fe^{3+}$ ($LaSrFeO_4$) standards is modest, but discernible at $\mu \sim 0.5$ [24]. The substantial broadening/degradation of the main peak between the $LaSrFeO_4$ and $SrFeO_{3-\delta}$ is also related to the $Fe^{3+}$ to $\sim Fe^{4+}$ crossover. The chemical shift between the $Fe^{3+}$ ($LaSrFeO_4$) and the $Fe^{2+}$ (FeO) spectra is substantial. The LaOFeAs spectrum is close to, but shifted to lower energy relative to the $Fe^{2+}$ (FeO) spectrum. It's consistent with essentially $Fe^{2+}$ $d^6$ character of ground state in LaOFeAs contrary to essentially covalent character $Fe^{2+}$ $d^7L + d^6$ (L is for a ligand hole) of the ground state in FeO [26,27].

The region in the ~10 eV below the main edge (see inset of Figure 1) in the Fe-K spectra is referred to as the pre-edge. The spectral features in the pre-edge region involve transitions into unoccupied final states of d-character either via quadrupole or dipole transitions [15-19, 24, 25]. In general the pre-edge features in Figure 1-inset can be seen to shift to lower energy with decreasing Fe-valence. The modest down-shift of the LaOFeAs pre-edge feature, relative to that of FeO, is again consistent with the somewhat higher average d-count in its electronic configuration. The pre-edge feature in LaOFeAs is particularly strong due to the fact that the local tetrahedral ligand field explicitly allows dipole transitions into 3d related states [25]. Here the other standard compounds have octahedral symmetry with weaker quadrapole transition pre-edge features. Thus the pre-edge feature of LaOFeAs is also consistent with its tetrahedral environment and a hybridized $Fe^{2+}$-centered electronic configuration with a somewhat larger average d-count than FeO.

In Figure 2 we compare the Fe K-edge XANES results on the $LaO_{1-x}F_xFeAs$ materials with x=0, 0.11 and 0.33. A small but clear downward chemical shift in the energy of the main edge of the x= 0.11 spectrum (relative to the x=0 spectrum) can be seen. This chemical shift is still larger in the x=0.33 specimen. We will follow Tranquada et al.'s [14] use of a difference absorption spectrum to identify the small spectral shifts associated with electron donation to Cu in the electron superconductor $Nd_{2-x}Ce_xCuO_4$ [14]. Accordingly we show in Figure 2 the difference spectra ($\Delta\mu$), obtained by subtracting the x=0 spectrum from the x>0 spectra. Four features (labeled 1-4 in the figure) are noted in the $\Delta\mu$ difference spectra shown in Figure 2.

The large $\Delta\mu$ spectral feature, at the energy labeled 1 in Figure 2, occurs at the very bottom of the main edge. In analogy with the interpretations of Tranquada et al. [14] we associate this feature with the doping induced enhancement of the antibonding-Fe-4p states associated with the $Fe^{1+}$ $d^7$ configuration, which would be expected to be enhanced by electron donation to the $Fe^{2+}$ $d^6$ site in the patent material.

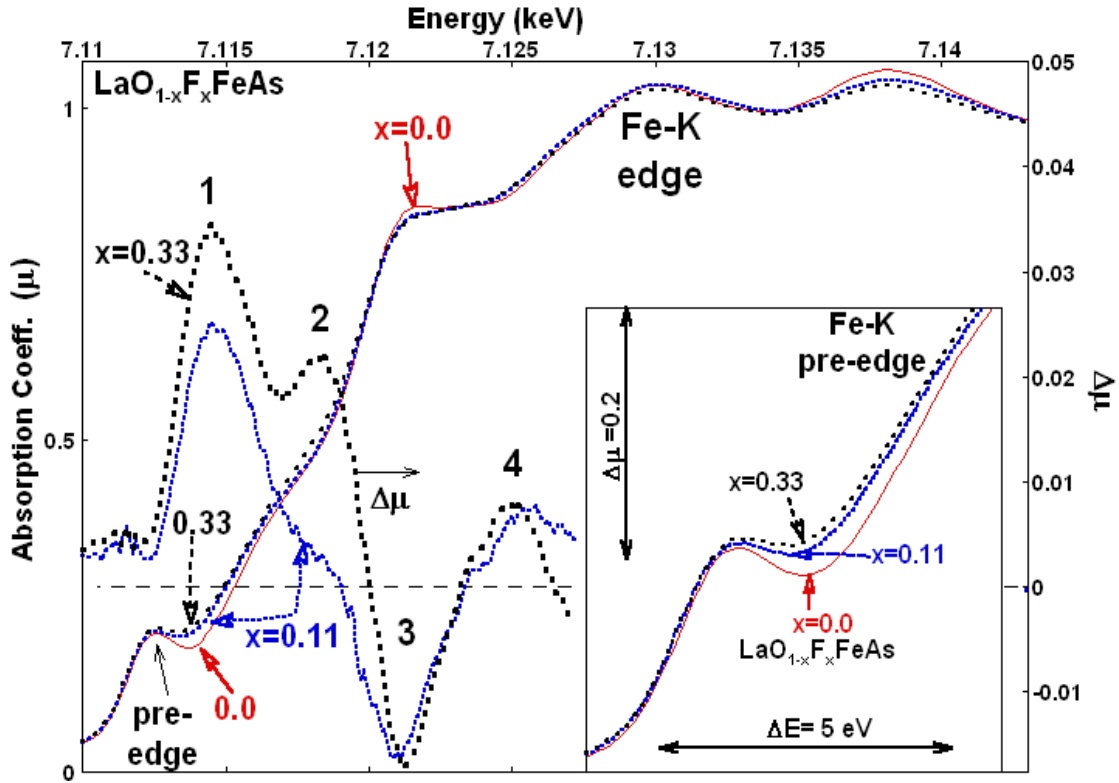

Figure 2: (color on line) Fe-K edge spectra for $LaO_{1-x}F_xFeAs$ with x=0 and 0.11 and 0.33. The difference spectra ($\Delta\mu$), obtained by subtracting the x=0 spectrum from the other spectra, are plotted on an expanded scale on the right. Inset: A comparison of the pre-edge region of these spectra.

This 1-feature constitutes the largest F-doping induced spectral change and grows uniformly from x=0 to x=0.11 to x=0.33. The second $\Delta\mu$ spectral feature, labeled 2 in the figure, is only a shoulder in the x= 0.11 spectrum but has grown into a resolved peak in the x=0.33 spectrum. The spectral shift to lower energy of the combined 1 and 2 features is consistent with a doping induced $Fe^{1+}$ admixture into the $Fe^{2+}$ parent material. Similar spectral changes have been observed in planar-nicklelate materials where chemical modifications have introduced $Ni^{1+}$ admixtures into $Ni^{2+}$ parent compounds [15]. The negative minimum in the $\Delta\mu$-curves near point 3 in Figure 2 is associated with the rounding/broadening of the sharp shoulder-feature in the x=0 spectrum in this energy range. Finally the positive $\Delta\mu$ curve peak, in the energy range labeled 4, is again an indicator of the chemical shift of the edge to lower energy accompanying the doping of charge to the Fe site.

Comparison of the x=0 and x>0 pre-edge feature regions in the inset of Figure 2 indicates that although the net spectral intensity increases with doping the pre-edge structure becomes less distinct. In order to compare the spectral weights of the 3d-state pre-edge features, spline fits connecting the "background regions" before-and-after the pre-edge have been subtracted from the each of the spectra [see for example reference 16]. The pre-edge only features so obtained are shown in the inset of Figure 3. The less-distinct nature of the doped pre-edge features can clearly be seen to be related to their loss of intensity relative to the x=0 pre-edge feature. Indeed in the inset of Figure 3 the x-dependence of the integrated pre-edge feature area (normalized to the x=0 feature area) can be seen to decrease systematically with the initial dopant concentration x.

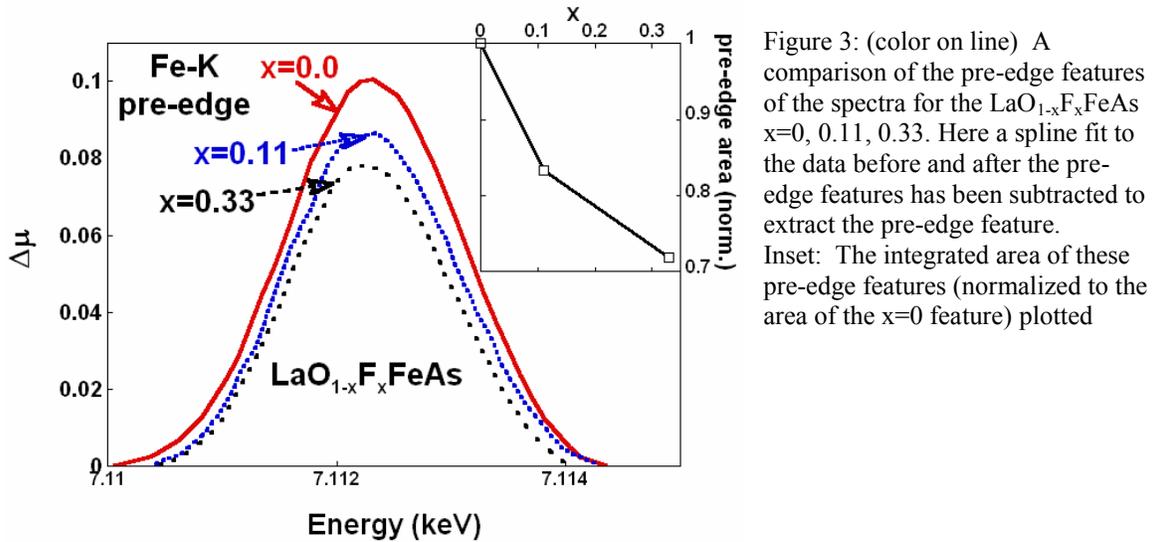

Figure 3: (color on line) A comparison of the pre-edge features of the spectra for the $LaO_{1-x}F_xFeAs$ x=0, 0.11, 0.33. Here a spline fit to the data before and after the pre-edge features has been subtracted to extract the pre-edge feature. Inset: The integrated area of these pre-edge features (normalized to the area of the x=0 feature) plotted

Within the accuracy of our experiment, pre-edge area decreases almost linearly with doping. It's consistent with $Fe^{2+}$ $d^6$ being converted to $Fe^1$ $d^7$ states. This interpretation agrees with results of the band structure calculations showing that the density of state (DOS) near $E_F$ is close to pure Fe $d$ in character [20, 22].

*3.2 As-K edge results*
     As $K$- edge results for the $LaO_{1-x}F_xFeAs$ specimens are shown in Figure 4. These high energy spectra are subject to substantial core hole and monochromator-resolution spectral broadening. Nevertheless three distinct features labeled A-B-C are clearly seen in Figure 4. The origins of the observed features in layered transition metal (T=Co, Cu) compounds with tetrahedral coordination to p-ligands (X=Si,Ge) are presented in the Appendix to avoid breaking the focus on the $LaO_{1-x}F_xFeAs$ systems in the main portion of the text.
The A-features at the onset of the As $K$-edge, related to the As(p)-Fe(d) hybridized states above $E_F$ is of primary interest. Similar at-edge-onset features in O K-edge spectra drew much attention in transition metal oxide compounds and Cu-based high $Tc$ superconductors [26,13]. The structure of the T-d density of states was clearly observed in A-features in binary T-O compounds [28]. In the cuprates, hole doping was accompanied by a concomitant A-feature appearance/enhancement of spectral weight that emphasized the importance of the O- charge carriers in vicinity of the Fermi level [13]. In other example S $K$- edge measurements across the metal-to-insulator (MI) transitions in $CuIr_2S_4$ [29] provided important insight into the electronic structure changes across the MI transition.

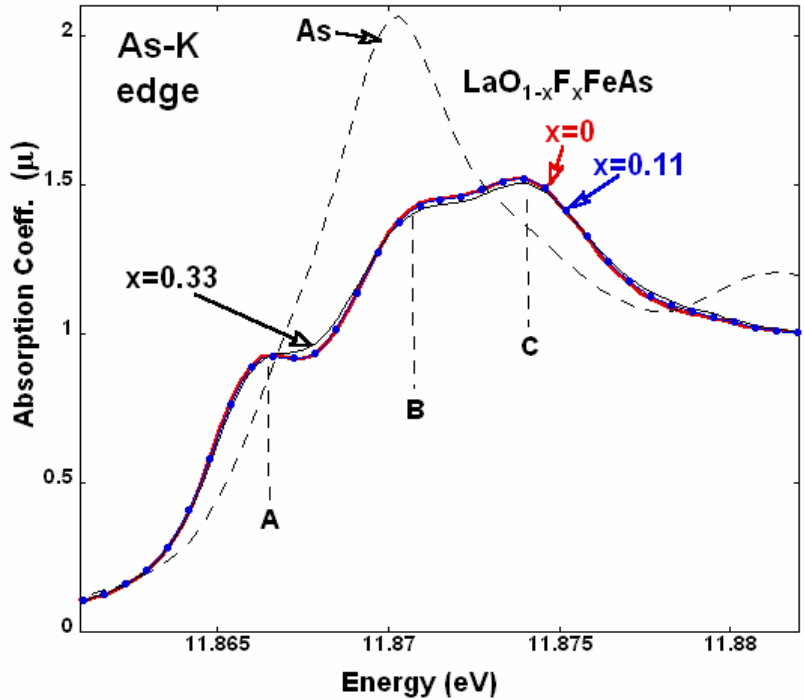

Figure 4 The As-K edge spectra for elemental-As, and $LaO_{1-x}F_xFeAs$ with x=0, 0.11 and 033.  Note the A-future associated with Fe-d/As-p hybridized states just above the Fermi level.

Overall the A-feature manifests essentially no change in position and intensity with the F-doping in this system. Within the limits of the experimental resolution, there seems to be little if any doping induced electron transfer to the As site.  This observation is consistent with the band structure calculations where the near-$E_F$ have essentially pure Fe-d character [20-22]. The states associated with the A-feature lie at least 1 eV above the Fermi level in the band structure calculations [20-22].   In contrast to the cuprates, the absence of the A- feature doping response emphasizes the lack of As-p character in the states near the Fermi level.

Regarding the two remaining features (B and C), comparing the As-*K* edges of $LaO_{1-x}F_xFeAs$ with x=0 and x=0.11 one notes no evidence for spectral changes within the limits of experimental uncertainty.  In the case of the x=0.33 an overall broadening of the B-C features is seen.  At present interpretation of this broadening in terms of electronic structure changes is not deemed warranted. Thus, within the intrinsic experimental broadening, our results do not manifest evidence for F-doping induced modifications in the As-site projected electronic structure. Higher resolution polarized XANES experiments are called for resolving potentially small changes.

### *3.3 La-$L_3$ edge Measurements*

Although the valence of La is not an issue in solids the XAS fine structure (FS) oscillations above the La $L_3$ edge can provide some insight into local structure and interatomic distances.  The sharp peak at the $L_3$ edge of La in LaOFeAs, see Figure 5, is due to intense, atomic like, 2p to 5d transitions and, for historical reasons, is referred to as a "white line" (WL) feature [30].  The smaller absorption coefficient oscillations above the edge are due to single and multiple scattering from the near neighbor atomic coordination shells around the absorbing atom, La in this case.  In transition metal and rare earth oxide compounds the most intense FS feature typically occurs in the 25-50 eV range above the 5d-WL feature and is sometimes referred to as a continuum resonance (CR) feature [30].

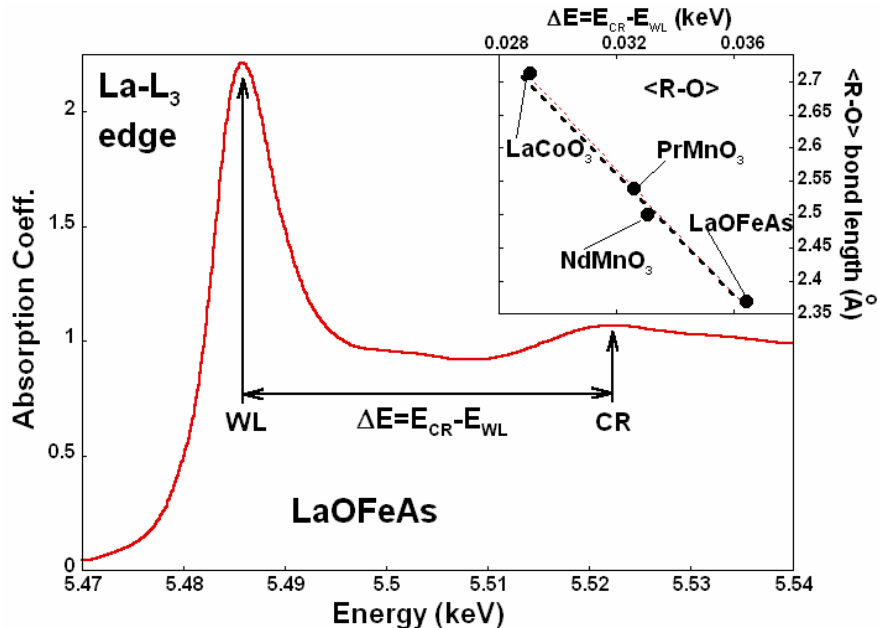

Figure 5.   La $L_3$- edge of LaOFeAs.  Note the "white line" (WL) and the continuum resonance (CR) features at the energies $E_{WL}$ and $E_{CR}$ respectively.  The energy difference $\Delta E = E_{CR}-E_{WL}$ is plotted versus the average first coordination shell rare earth (R), oxygen bond length, <R-O> in the inset of the figure.

The energy of this CR feature peak, relative to the WL energy, $\Delta E = E_{CR}-E_{WL}$ has been empirically found to scale with the average radius of the first ligand shell (in this case the O-shell) [30-32].  The containing potential of the filled shell ligand atoms, EXAFS type based arguments, or Natoli's rule, have been invoked for the $\Delta E$ dependence upon the the average ligand shell radius [30-32].  For rare earth (R) compounds with modestly different average first shell average distances, the bond length <R-O> varies approximately linearly with $\Delta E$.  Despite substantial excursions of the individual R-O distances about the average, the comparison of the $\Delta E$ in standards typically provides a good first estimate of <R-O>.  In Figure 5 we plot the known <R-O> values versus the $\Delta E$ for a series of rare earth compounds studied by our lab.  The measured value $\Delta E$ = 36.4 eV and the <La-O> value of 2.37 Å, extracted from the structure results of Nomura et al. [33], has been included in this plot.  The LaOFeAs compound data point agrees quite well with a linear correlation.  Although this observation is simply verifying the utility of an empirical relation for these materials it offers a quick, local probe of the [RO] layer interatomic distanced in this class of compounds.

### IV. Conclusions

Our Fe $K$-edge XANES measurements clearly demonstrate that F substitution in the LaO$_{1-x}$F$_x$FeAs system donates electronic charge to the Fe sites and that this charge has Fe-d character. The As *K*- edge measurements show that the As-site projected electronic structure is not responsive to the F doping (only a spectral broadening at the highest F doping level was observed). Both Fe and As *K*-edge measurements indicate that F-doping transfers charge into nearly pure Fe-d states at $E_F$, consistent with the band structure calculations [20-22]. Within the cluster model nomenclature these results indicate little involvement of As-ligand hole in the electronic configuration of these materials. This simple observation is important in its dramatic contrast to the high-$T_C$ cuprates where the charge superconducting carriers are strong Cu-d/O-p admixtures. Namely the carriers involved in the superconductivity and these new materials would appear to be solely F-d state related. Finally the fine structure XAS results just above the La-$L_3$ edge yield an estimate of the La-O interatomic distance consistent with the literature and with bond lengths in other rare earth/oxide containing compounds.

**Appendix: X-K edges in layered tetrahedral T-X compounds**

To discuss the origin of the features A, B, and C on the As *K*-edge of the LaO$_{1-x}$F$_x$FeAs we consider here selected XANES results for p-block elements (X) in related transition metal (T) T-X compounds. {LaO}[FeAs], and {Th}[Cr$_2$Si$_2$] structure materials are examples of a class of layered compounds based upon the building block of strongly bonded T-X planes with tetrahedral T coordination to the p-block element X [1-11, 34]. The T-X layers are separated by a second type of plane with a different type of bonding; the ionically bound {LaO} plane in the former, and the metallically bound {Th} plane in the latter. The interlayer bonding is typically ionic, e.g. {LaO}$^{+1}$[FeAs]$^{-1}$ and {Th}$^{+\alpha}$[Cr$_2$Si$_2$]$^{-\alpha}$. Here some of the Th$^{4+}$ electrons are involved in intra-layer metallic bonding so that a reduced number, $\alpha$, contribute to the interlayer ionic boding. As pointed out by Hoffman and coworkers [34], intra-layer direct T-T overlap bonding is important in these ThCr$_2$Si$_2$-structure (122) materials. They also noted that X-X bonding bridging the {Th}-layer can, in some cases, be important [32].

In Figure A1 and we show X-K near edge XANES results for X=Ge and Si in the R$^{3+}$T$_2$X$_2$ compounds with T= Cu and Co. The covalent nature of the T-X bond in ThCr$_2$Si$_2$ structure compounds is most apparent from the Si-K edge spectra in Figure A1a. The A-feature immediately above the Fermi-level/edge-onset is due to Si-p hybridized with empty T-d states. This interpretation is consistent with extensive binary T-X compound Si-*K* edge measurements. [35]. The A-feature can be seen to increases in intensity and move down in energy between the T= Cu to Co spectra of R$^{3+}$T$_2$Si$_2$. The increase of its spectral weight is due to higher number of d-holes in Co-based compound. Increased cohesive energy (from T-T metallic and covalent T-X bonding) contributes to a lowering of the relative Fermi energies between the compounds.

Similar effects, albeit more subtle, are visible in the Ge-K edge spectra of the T= Cu and Co, R$^{3+}$T$_2$Ge$_2$ compounds in Figure A1b. Specifically the A-feature increases in intensity and moves down in energy between the T= Cu and Co spectra. Again these changes reflect the d-state electronic structure changes with the increase in the number of d-holes, and the increase of the cohesive energy between the Cu and Co materials. In the case of the Fe-As compounds, discussed in the text, a stronger Fe to As charge transfer presumably occurs giving them a stronger ionic character compared to the Si and Ge compounds discussed in this appendix. Nevertheless the hybridized Fe-As states above the Fermi energy should lead to similar structures at the *K* absorption edge.

The Ge-K edge spectra in Figure A1b show two additional features labeled B, and C. The highest energy feature (labeled C in Figure 3b) moves to higher energy between the Cu and Co materials. This feature is presumably the highest T-X anti-bonding orbital and its movement to higher energy is consistent with the increase in the T-X bonding/anti-bonding splitting on going from X=Cu to Co. A similar high energy antibonding feature has been seen in the $T(4d)$-$L_{2,3}$ edges of 122 compounds and has been shown to track the expected T-X bonding strength in a large group of such materials [36]. The intermediate B-feature is due to in-plane vs. out-of-plain split in the T-X anti-bonding orbitals and is relatively insensitive to the charge transfer.

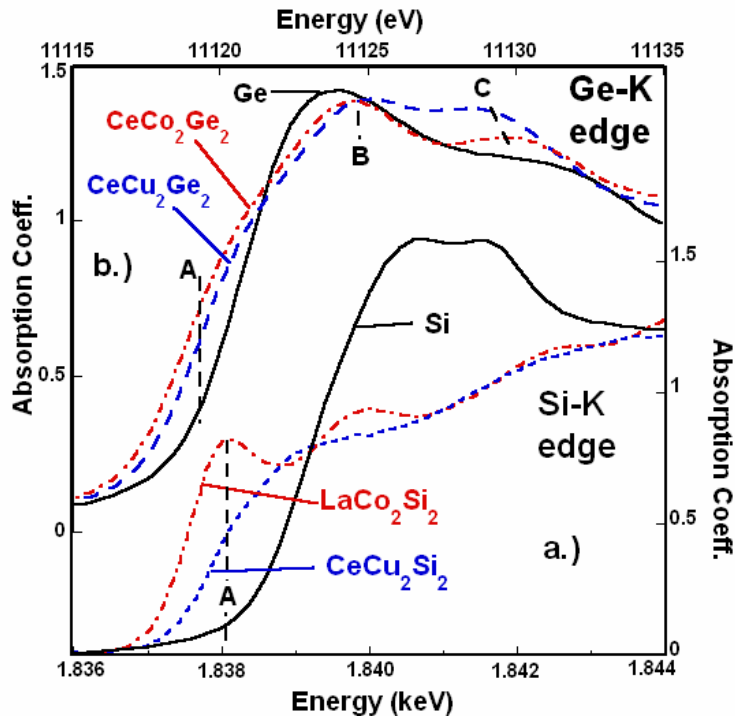

Figure A1 (a). The Si-K edge spectra for elemental-Si, $R^{3+}T_2Si_2$ (R=La and Ce, and T=Cu and Co). Note the A-feature at the edge onset which is associated with T-3d states.

(b) The Ge-K edge spectra for elemental-Ge, $R^{3+}T_2Ge_2$ (R=Ce, and T=Cu and Co).

**Acknowledgements**

This work was supported in part by the National Science Foundation under MRSEC NSF-DMR- 0520471 and the U. S. Department of Energy Office of Basic Energy Sciences (DOE-BES) under Grant DE-FG02-07ER46402.